\documentclass[acmsmall,screen]{acmart}
\usepackage{booktabs}
\usepackage{tabularx}
\usepackage{url}
\usepackage{makecell}
\usepackage{array}
\usepackage{multirow}
\usepackage{booktabs}

\usepackage{amssymb} %
\newcolumntype{N}{@{}c@{}} 

\usepackage{ragged2e}
\usepackage{natbib}
\usepackage{url}
\usepackage{hyperref}
\usepackage{subfigure}

\newcolumntype{L}[1]{>{\RaggedRight\hspace{0pt}}p{#1}}

\AtBeginDocument{%
  }

\begin{document}

\title{Device-Level Optimization Techniques for Solid-State Drives: A Survey}

\author{TIANYU REN}
\affiliation{%
  \institution{Tsinghua University}
  \city{Beijing}
  \country{China}}
\email{m0llyren.dyc@gmail.com}

\author{YAJUAN DU}
\affiliation{%
  \institution{Wuhan University of Technology}
  \city{Wuhan}
 \state{Hubei}
 \country{China}}
 
\author{JINHUA CUI}
\affiliation{%
 \institution{Huazhong University of Science and Technology}
 \city{Wuhan}
 \state{Hubei}
 \country{China}}

\author{YINA LV}
\affiliation{%
  \institution{Xiamen University}
  \city{Xiamen}
  \state{Fujian}
  \country{China}}

\author{QIAO LI}
\affiliation{%
  \institution{Mohamed bin Zayed University of Artificial Intelligence}
  \city{Abu Dhabi}
  \country{The United Arab Emirates}
 }
\email{qiaoli045@gmail.com}

\author{CHUN JASON XUE}
\affiliation{%
 \institution{Mohamed bin Zayed University of Artificial Intelligence}
  \city{Abu Dhabi}
  \country{The United Arab Emirates}}

\renewcommand{\shortauthors}{Ren et al.}

\begin{abstract}
Solid-state drives (SSDs) have revolutionized data storage with their high performance, energy efficiency, and reliability. However, as storage demands grow, SSDs face critical challenges in scalability, endurance, latency, and security. This survey provides a comprehensive analysis of SSD architecture, key challenges, and device-level optimization techniques. We first examine the fundamental components of SSDs, including NAND flash memory structures, SSD controller functionalities (e.g., address mapping, garbage collection, wear leveling), and host interface protocols. Next, we discuss major challenges such as reliability degradation, endurance limitations, latency variations, and security threats. We then explore advanced optimization techniques, including error correction mechanisms, flash translation layer (FTL) enhancements, and emerging architectures like zoned namespace (ZNS) SSDs and flexible data placement (FDP). Finally, we highlight open research challenges, such as QLC/PLC NAND scalability, performance-reliability trade-offs, and SSD optimizations for AI/LLM workloads. This survey aims to guide future research in the development of next-generation SSDs that balance performance, endurance, and security in evolving storage ecosystems.
\end{abstract}

\keywords{Solid-State Drives, Device-Level Optimization, NAND Flash Memory}

\maketitle

\section{Introduction}\label{sec:intro}
NAND flash-based SSDs have become the cornerstone of modern storage, offering superior performance, energy efficiency, and reliability over HDDs \cite{7906530}. Their rapid evolution is driven by growing demands from cloud computing, big data, AI, and high-performance computing \cite{FAST2022operational}. However, widespread adoption in enterprise and consumer markets reveals significant challenges in reliability, endurance, performance consistency, and security, necessitating device-level innovations \cite{KimCM19,LuoLLS23}.
Modern SSD architecture comprises several performance-defining components. The foundation is NAND flash memory, which has evolved from Single-Level Cell (SLC) and Multi-Level Cell (MLC) to Triple-Level Cell (TLC), Quad-Level Cell (QLC), and Penta-Level Cell (PLC) technologies \cite{234980}, increasing storage density and reducing cost per bit but introducing challenges in write endurance, read latency, and error rates \cite{9251942}. The SSD controller executes critical functions via Flash Translation Layer (FTL) algorithms, managing address mapping, garbage collection, and wear leveling. These components interface with hosts through protocols like Serial ATA (SATA) and Non-Volatile Memory Express (NVMe), the latter dominating high-performance applications due to its low latency and high parallelism \cite{SATAIO,NVMe}.

Despite their advantages, SSDs face several inherent limitations distinct from traditional storage. The write-once nature of NAND flash requires complex garbage collection, leading to write amplification and unpredictable latency \cite{wang2022separating}. Reliability concerns arise from program/erase (P/E) cycling effects, read disturb phenomena, and data retention issues that intensify with advanced NAND technologies \cite{2591312}. Additionally, flash characteristics introduce security vulnerabilities including data remanence attacks and specialized malware like ransomware that exploit SSD wear-out mechanisms \cite{tripathy_FormalModelingVerification_2023,meijer2019self}.

Recent years have witnessed significant advances in SSD device-level optimization. Advanced error correction coding schemes, particularly low-density parity-check (LDPC) codes, have become essential for maintaining data integrity in high-density NAND flash \cite{zhao2013ldpc}. Novel FTL designs address garbage collection bottlenecks through techniques like BER \cite{9104673} and GuardedErase \cite{277832}. 
The storage industry has also developed new SSD architectures, including Zoned Namespace (ZNS) SSDs and Flexible Data Placement (FDP) designs that enable host-assisted performance optimization \cite{HotStorage2020lsmGCforZNS,allison2025towards}. 
These innovations bridge the growing gap between application requirements and raw NAND flash characteristics.

This survey paper provides a comprehensive examination of SSD device-level optimization techniques, structured as follows: Section \ref{sec:background} discusses SSD architecture and key components, including NAND flash fundamentals, controller functions, and interface protocols. 
Section \ref{sec:challenges} analyzes the major challenges facing SSDs in terms of reliability, endurance, latency, and security. 
Section \ref{sec:survey} reviews state-of-the-art optimization techniques across multiple dimensions of SSD design. 
Finally, Section \ref{sec:open} discusses open research challenges and future directions, particularly focusing on emerging NAND technologies and novel application workloads.

The contributions of this survey are threefold: 
(1) a systematic organization and analysis of contemporary SSD optimization techniques; 
(2) identification of key challenges and trade-offs in SSD design; 
and (3) a forward-looking perspective on emerging research directions. 
By synthesizing knowledge from academic research and industry practice, this survey paper aims to serve as a valuable reference for researchers and engineers working on next-generation storage systems. 
As SSDs continue to evolve to meet the demands of increasingly data-intensive applications, understanding these device-level optimization techniques becomes crucial for pushing the boundaries of storage performance, efficiency, and reliability.

\section{Related Work}\label{sec:related}
From 2009 to 2025, 22 survey papers have analyzed SSD technologies across six key dimensions: FTL-layer optimizations \cite{CHUNG2009332,LUO2021480,JIN20191,DBLP,TRIPATHY2022102334}, underlying chip enhancements \cite{9249760,WEI2020113738,articlerp3d}, architectural innovations \cite{AlsalibiMAS18,8283744,101007s00778019005598,doekemeijer2022keyvaluestoresflashstorage,articleocssd,10.11453708995,1011453708992}, scenario-specific optimizations \cite{9265320,tehrany2023surveyintegrationnandflash,10606974,10.11453723167}, lifetime evaluation metrics \cite{9040759}, and simulator development \cite{gheibi2025survey}.
However, existing surveys exhibit clear limitations. FTL-layer optimization surveys \cite{DBLP,LUO2021480} focus only on address mapping techniques and are limited to literature published in 2020.
Chip optimization reviews \cite{9249760,WEI2020113738,articlerp3d} fail to address emerging AI-scenario advancements. Architectural studies \cite{AlsalibiMAS18,8283744,101007s00778019005598,doekemeijer2022keyvaluestoresflashstorage,articleocssd,10.11453708995,1011453708992} lack a systematic analytical framework. Scenario-specific surveys \cite{9265320,tehrany2023surveyintegrationnandflash,10606974,10.11453723167} present incomplete technology roadmaps. Research on lifetime metrics \cite{9040759} overlooks recent solutions addressing capacity-driven degradation, while the simulator review \cite{gheibi2025survey} overemphasizes basic implementation mechanisms and ignores cutting-edge breakthroughs.
In contrast, this survey adopts a device-side perspective, unifying SSD's four core challenges (reliability, endurance, performance, and capacity). It provides a comprehensive analysis of FTL algorithms, architectural evolution, and AI-driven optimizations. By extensively incorporating newly published literature from 2020 to 2025, the survey effectively bridges critical gaps in both timeliness and comprehensiveness found in existing research.

\begin{figure*}[htbp] 
\vspace{-0.2in}
\centering
\begin{tabular}{l}
\subfigure[The Floating Gate Cell Structure.]{\includegraphics[width=0.5\columnwidth,trim=0 0 0 0,clip]{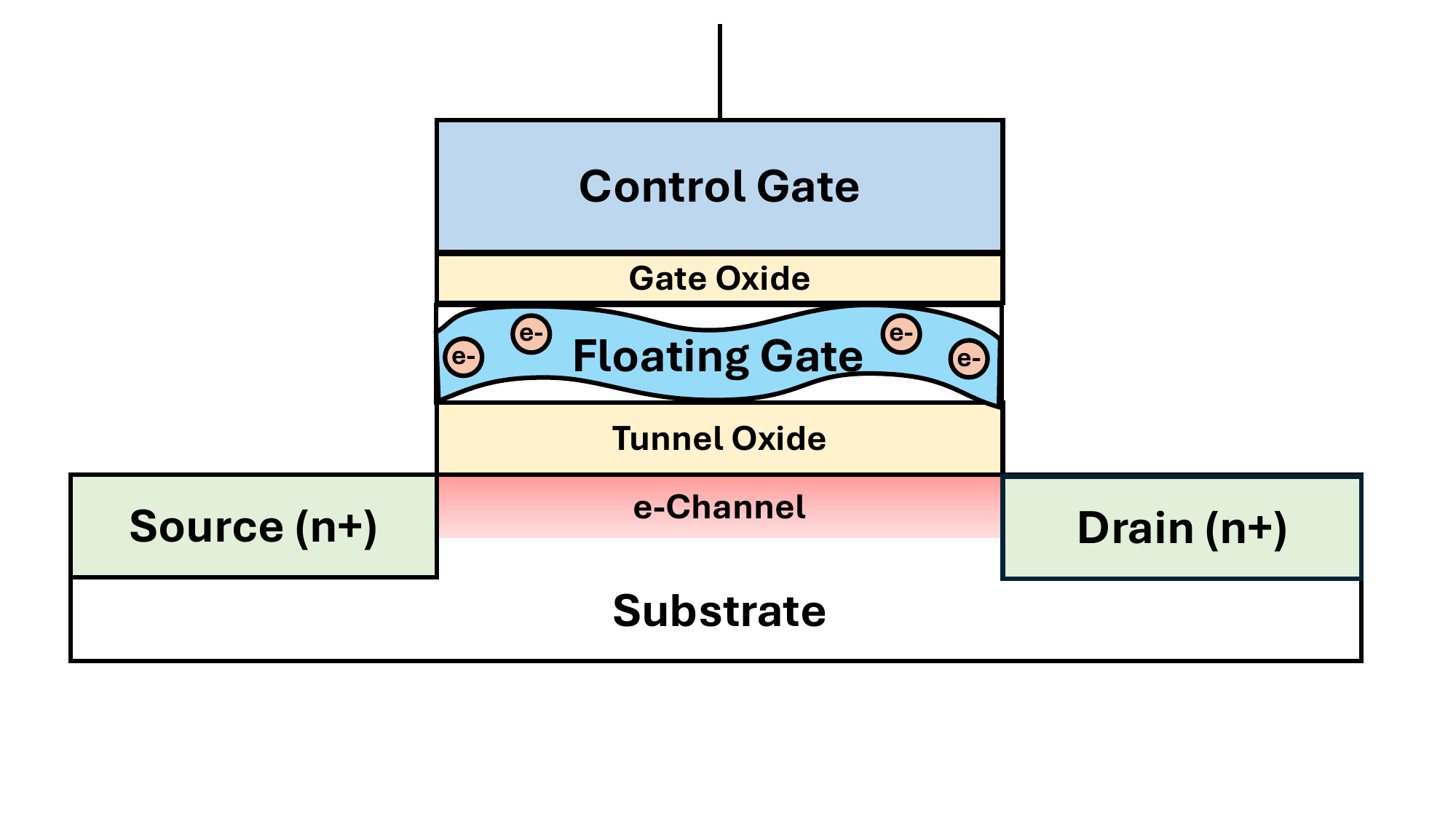}}
\subfigure[The Charge Trap Cell Structure.]{\includegraphics[width=0.5\columnwidth,trim=0 0 0 0,clip]{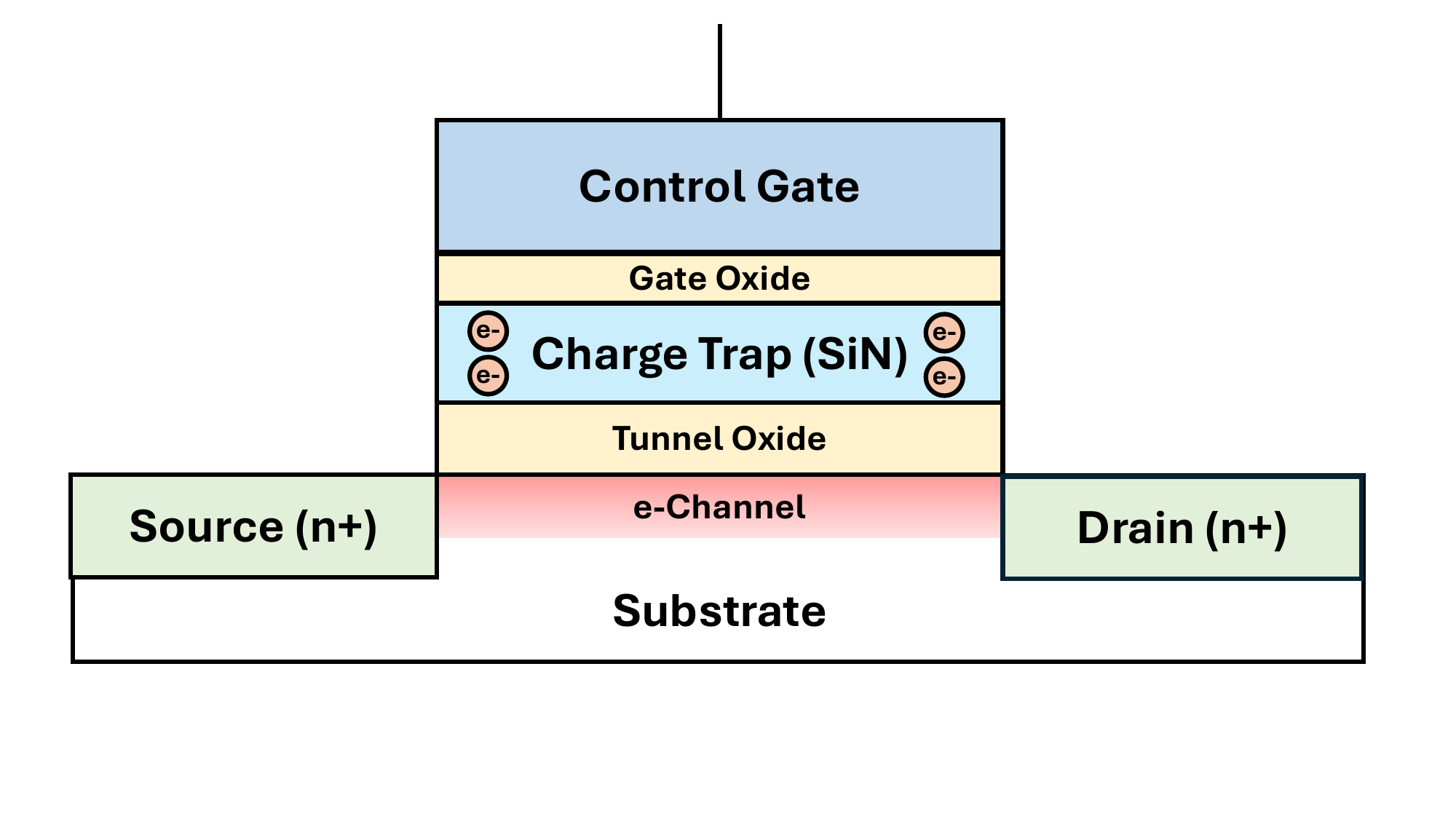}}
\end{tabular}
\vspace{-0.2in}
\caption{The Comparison between CT and FG Cell Structure }\label{fig:cell}
\vspace{-0.3in}
\end{figure*}

\section{SSD Architecture and Key Components}
\label{sec:background}
This section provides a comprehensive introduction to SSD architectures, giving a holistic understanding of the entire SSD structure, from the basics of NAND flash memory cells (including floating gate and charge trap) to the sophisticated controller algorithms that manage data flow. The introduction also extends through critical components, including DRAM caching mechanisms, buffer management strategies, and host interface protocols (SATA, NVMe, PCIe) that define communication standards, providing complete fundamental insight into how these integrated systems optimize SSD performance, reliability, and endurance from the perspective of the device level.

\subsection{NAND Flash Memory Fundamentals}
This subsection introduces the foundational technology of NAND flash memory from its physical architecture and operational principles. We detail how memory cells are organized into hierarchical structures, including pages, blocks, and planes that define the physical framework of flash memory. The discussion further covers fundamental operations: program, erase, and read operations.

\subsubsection{\textbf{Flash Memory Cell Basics}}

Toshiba introduced the first flash memory chip (256 Kb) \cite{masuoka1985256k} and presented Fowler-Nordheim (FN) tunneling in 1990 \cite{shirota19902}. Samsung achieved the first mass-produced chip (32 Mb) in 1995 \cite{suh19953} and pioneered MLC technology (128 Mb) in 1996 \cite{488501}. During the 2D era, SanDisk announced the first mass-produced TLC chip (16 Gb) in 2008 \cite{cernea200834mb}. A major shift occurred in 2014 when Samsung launched the first 3D V-NAND with 24-layer stacking \cite{park2014threeF}, followed by the world's first 1Tb QLC V-NAND in 2018 \cite{lee20181tb}. In 2023, Intel proposed a 1.67TB, 192-layer PLC flash chip \cite{khakifirooz20231}, marking continued density advancement.

NAND flash memory is based on MOS transistors (cells) where threshold voltage variations represent stored data states \cite{DBLP}. Storage cells utilize two primary technologies: floating gate (FG) \cite{kahng1967floating} and charge trap (CT) \cite{kim2006future}. Both share a core structure comprising a control gate, dielectric layers, tunnel oxide, and source/drain electrodes. The key distinction lies in charge storage: FG employs a conductive floating gate for free electron movement, while CT confines charges in a silicon nitride trapping layer, restricting electron mobility. CT technology demonstrates superior resistance to inter-cell interference.
In current generations, Solidigm and Micron adopt FG architecture in products like Solidigm's 192-layer QLC SSD \cite{d5p5} and Micron's mobile TLC 3D NAND \cite{Micronmobile}. Conversely, Samsung and Kioxia implement CT designs such as Samsung's PM1743 QLC SSDs \cite{samsungct} and Kioxia BiCS flash \cite{Kioxiact}.

\subsubsection{\textbf{NAND Flash Structure}}

NAND flash has evolved from 2D planar to 3D vertical stacking \cite{yoon2022fundamentals}. In 2D architecture, memory cells form series-connected strings controlled by source and drain select transistors, with shared bitline contacts and parallel wordlines defining logical pages. Density per wordline is governed by individual cell storage capability. The transition to 3D NAND reorients the structure vertically, forming channels surrounded by stacked wordline layers for multi-tier cell integration. A 3D NAND device contains hundreds of blocks organized in 3D arrays, where cells along rows form wordlines, vertical alignments create strings, and horizontal wordline connections build layers, interconnected via bitlines and source lines. This progression continuously increases storage density, driven by cell architecture innovations from SLC (1 bit/cell) to MLC, TLC, QLC, and PLC (5 bits/cell) \cite{khakifirooz20231}.

\subsubsection{\textbf{Basic Flash Operations}}
In NAND flash memory, both read and program operations execute at the granularity of a page, while erase operations fundamentally occur at the block level. 

During \textbf{read operations}, the target page's wordline is biased at a relatively low read reference voltage, while other wordlines receive pass-through voltage ($V_{\text{PASS}}$). The bitline associated with the string is precharged to a specific sensing voltage. If the target cell is in an erased state, the corresponding sense amplifier detects current flow through the bitline. In contrast, no current is detected for programmed cells. 
\textbf{Programming operation} leverages the tunneling effect where electrons traverse the tunnel oxide under a strong electric field. The selected worldine receives program voltage, while the unselected wordlines are biased at $V_{\text{PASS}}$. The drain selector gate is driven to Voltage Drain Drain, and the source selector gate is grounded. This configuration creates an electric field that enables electron tunneling, trapping electrons in the storage layer, thus completing the programming of individual cells to a logical ``0'' state~\cite{gheibi2025survey}.
\textbf{Erase operations} apply 0 V to all wordlines, while the substrate receives a high erase voltage. 
The resulting electric field triggers electron tunneling out of the storage layer, and setting cells to a logical ``1'' state.

\begin{figure}[t]
\vspace{0.4cm}
\centerline{\includegraphics[width=0.9\columnwidth,trim=40 0 0 60]{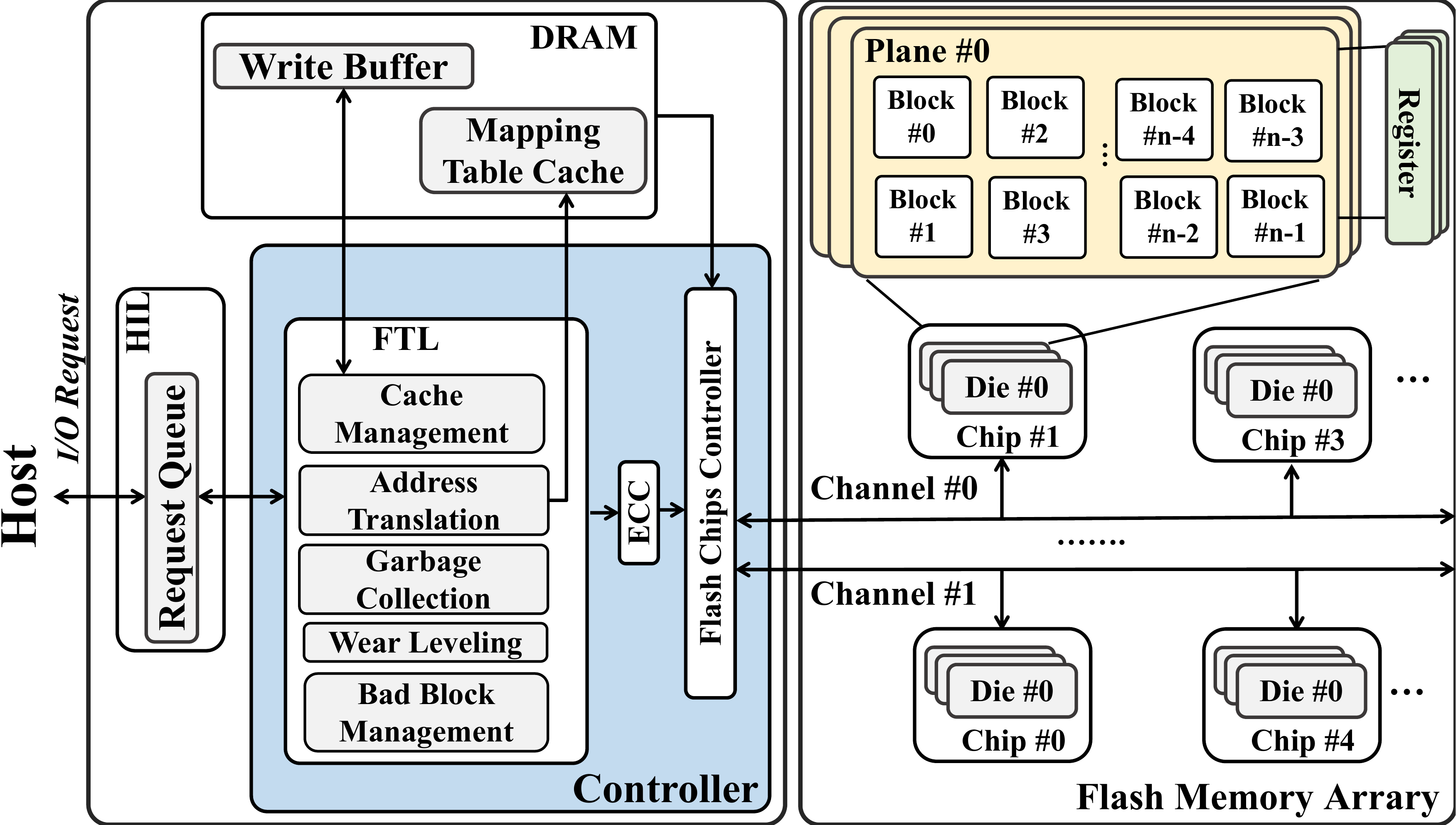}}
\vspace{-0.2cm}
\caption{The overview of SSD structure.}
\label{fig:ssd}
\vspace{-0.5cm}
\end{figure}

\subsection{SSD Controller}
The SSD controller is the core processing unit of an SSD, primarily responsible for handling commands and data from the host, as well as managing flash memory chips.
Its key components include the host–interface logic (HIL), flash translation layer (FTL), on-board DRAM, and flash chip controllers~\cite{malladi2017flashstoragesim}, as shown in Fig.~\ref{fig:ssd}.
The FTL is the core SSD firmware algorithm responsible for performance, reliability, and endurance. 
Its key functions include address mapping, garbage collection, wear leveling, error correction codes (ECC), and bad block management. 
Controllers further improve reliability by integrating data compression and decompression capabilities~\cite{gheibi2025survey}.

\subsubsection{\textbf{Address Mapping}}
Address mapping establishes a one-to-one correspondence between host logical addresses and physical data locations, maintained in a table loaded in DRAM during initialization. There are three main address mapping methods.
\textbf{Block mapping} operates at block granularity, mapping logical blocks to arbitrary physical blocks while keeping intra-block page offsets fixed \cite{DBLP}. This minimizes table size but incurs high overhead: sub-page updates require reading, modifying, and rewriting the entire block.
\textbf{Page mapping} uses page-level granularity, enabling flexible one-to-one logical-to-physical page translation \cite{ChungPPLLS06}. As the dominant method, it offers superior I/O performance but demands a larger mapping table due to higher page-to-block ratios.
\textbf{Hybrid mapping} combines both approaches by segmenting logical addresses into blocks while allowing page-level mapping within blocks \cite{LiuSLZQZLZ25}. This balances performance and table size between pure block and page mapping.

\subsubsection{\textbf{Garbage Collection}}
Garbage collection reclaims storage space when free capacity falls below a set threshold. 
A common strategy like the greedy algorithm selects the block with the most invalid pages for erasure, thus efficiently creating new writable space.
The selected block is termed the victim block.
Before erasing this block, all valid pages within it must be read out and copied to a free block.
In open-channel SSDs (OCSSDs), valid pages are read from the OCSSD to the DRAM and then are migrated to the OCSSD~\cite{LiHTSBG18,picoli2020open}.
When garbage collection migrates valid data pages from the victim block to the newly free block, it inherently risks generating uneven wear patterns across the NAND flash memory cells and a write amplification issue~\cite{desnoyers2012analytic}. 

\subsubsection{\textbf{Wear Leveling}}
Wear leveling maximizes SSD lifetime and reliability by evenly distributing write/erase operations across blocks, countering uneven wear from host update patterns and garbage collection. Wear leveling strategies are categorized as dynamic or static.
\textbf{Dynamic Wear Leveling} directs frequently updated hot data to blocks with low program/erase (P/E) cycles. This concentrates invalid pages in less-worn blocks, avoiding garbage collection on highly-cycled blocks \cite{ChangK02}.
\textbf{Static Wear Leveling} migrates infrequently updated data from low-P/E cycle blocks to higher-P/E cycle blocks. This reduces the likelihood of selecting heavily-worn blocks for garbage collection, freeing low-cycle blocks for frequently updated data \cite{MuruganD11,WangW12}.

\subsubsection{\textbf{Error Correction Codes}}
Due to factors such as data retention, program disturb, read disturb, manufacturing defects, and universal radiation, NAND flash memory cells may experience random bit errors, known as raw bit errors.
These accumulated errors can lead to data corruption or even system failure. 

Error Correction Code (ECC) adds redundant parity information to the original data, enabling the detection and correction of raw bit errors during read operations. 
When writing a page of data, the SSD controller splits it into smaller blocks. 
For each block, an ECC encoder generates a correction code, which is combined with the data block to form a complete codeword—stored together in the NAND flash physical page \cite{li2019ldpc}.
The level of protection is determined by the code rate, the ratio of data size to codeword size \cite{10.1145/3663478}. 
A higher code rate reduces storage overhead but weakens error correction capability, representing a key reliability-storage trade-off in SSD design.
During a read operation, the SSD controller retrieves the codeword from NAND flash. If errors are detected, the ECC engine attempts correction using the stored redundant bits. Should errors exceed the algorithm's correction capability, the controller adjusts the read voltage and retries. If all retries fail, an uncorrectable error is reported.
The two primary ECC algorithms are BCH (Bose-Chaudhuri-Hocquenghem)~\cite{lee20126}, which correct multiple errors in a single read operation, and LDPC (low-density parity-check)~\cite{mackay1996near}. LDPC codes utilize probabilistic information from multiple read operations to deliver stronger error correction capability, though this comes at the cost of higher decoding latency and storage overhead.

\subsubsection{\textbf{Bad Block Management}}
Bad block management (BBM)ensures SSD availability by detecting, isolating, and replacing failed NAND flash blocks. It maintains a bad block table (BBT) to record defective blocks, which arise from factory defects or wear-out—and skips them during writes~\cite{MicronBBM}.
Using spare blocks from the over-provisioning space, BBM dynamically maps replacements for failed blocks and maintains a dedicated mapping table to track these substitutions.

\subsection{DRAM Cache and Data Buffer}
This section outlines the architecture of request queues for I/O management, examines I/O scheduling strategies for request prioritization, analyzes buffer management for efficient caching.

\subsubsection{\textbf{Request Queue Design}}
SATA SSDs primarily use native command queuing to schedule multiple I/O requests that are sent from the host, which are first stored in a single request queue within the HIL, typically with a queue depth between 32 and 256, before being processed by the SSD~\cite{chen2019delay}.
For NVMe SSDs, I/O requests sent from the host over the PCIe bus are directly inserted into submission queues (SQs) within the HIL, where they await SSD processing. 
The completed requests processed by the SSD are then inserted into completion queues (CQs) within the HIL~\cite{FAST2018MQsim}. 
In addition, NVMe SSDs support multiple request queues, up to 64KiB pairs of I/O SQs and CQs~\cite{NVMe}. 

\subsubsection{ \textbf{I/O Scheduling Strategies}}
The I/O scheduling in HIL is primarily based on the queue protocol defined by the host, typically employing a first-in-first-out strategy within the same queue to arrange requests in chronological order~\cite{gouk2018amber}. 
Many host-interface protocols allow the host to assign different priorities to each request, where requests with higher priority are first scheduled. 
In NVMe SSDs with multiple queues, a weighted round-robin arbiter selects requests for processing, often prioritizing reads to boost performance~\cite{ISCA2018FLIN,9427266,10818146}.
The HIL splits the selected request into page-sized sub-requests, which are cached in the on-board DRAM. If free buffer slots are available, the data is inserted for immediate completion~\cite{yoo2013vssim,LiHTSBG18,JungZAKSSKK18,liu2022improving}.

\subsubsection{\textbf{Buffer Management Strategy}}
The DRAM controller, integrated with the SSD's CPU cores, manages cached requests using least recently used (LRU) or random replacement algorithms. It flushes the least recently used data to NAND flash when space is low or upon a host flush command to free up space for new requests~\cite{peng2024scalacache}.
Requests are allocated to a target page based on the SSD's parallelism and the idle state of its channel or chip.
Different allocation strategies, such as channel-first, die-first, plane-first, and chip-first, are used to optimize latency and throughput performance~\cite{FAST2018MQsim}.
In addition to caching data, DRAM also stores metadata and address mapping information.
DRAM capacity typically comprises about 0.1\% of total SSD storage with caching methods include set-associative and log-structured designs~\cite{allison2025towards}. 
The set-associative approach reduces tracking overhead for small objects but causes frequent random writes~\cite{mcallister2021kangaroo}, whereas log-structured caching generates flash-friendly sequential writes via infrequent batch updates~\cite{berg2020cachelib}.

\subsection{Host Interface Protocols}
The SSD host interfaces define the physical, link and protocol layers for host-device communication. Early SSDs adopted HDD-designed interfaces like SATA and SAS for easy integration. While SATA became the consumer-grade mainstream, its AHCI-based design limited parallelism; SAS offered enterprise reliability, but constrained flash potential.
To fully exploit flash parallelism and low latency, NVMe was developed. Operating directly over PCIe, it supports thousands of queues, reduces latency, and significantly improves throughput/IOPS, establishing itself as the high-performance standard. The following subsections detail these protocols.

\subsubsection{\textbf{Serial Advanced Technology Attachment (SATA)}} SATA serves as a widely adopted consumer-grade SSD interface utilizing the AHCI protocol originally designed for HDDs, favoring sequential access over flash parallelism with half-duplex operation constraining throughput potential \cite{2015nvmeperformance}, while its evolutionary revisions introduced NCQ optimization in SATA 2.0 and achieved 600 MB/s bandwidth in SATA 3.0 that remains bottlenecked by a single 32-command queue architecture limiting parallelism and queuing efficiency \cite{SATAIO}, making these SSDs suitable for consumer devices and cold data storage but inadequate for enterprise applications requiring hot-swapping or dual-port redundancy due to lacking high-availability features.

\subsubsection{\textbf{Serial Attached SCSI (SAS)}}
SAS is an enterprise storage interface designed for high-throughput computing with advanced control capabilities \cite{SAS2021reliability}. It employs three core protocols: SSP for host-device communication, SMP for expander management and large-scale connectivity, and STP for SATA device integration. Operating in full-duplex mode with dual-port support, SAS achieves speeds up to 48 Gb/s with deep command queues (128-256 commands), supporting thousands of drives in scalable direct-attached storage configurations.
Despite its scalability and reliability strengths, SAS's SCSI-based architecture predates modern flash technology, creating performance bottlenecks under concurrent workloads. Consequently, it cannot meet the low-latency requirements of contemporary data-intensive applications \cite{FAST2022operational}.

\subsubsection{\textbf{Non-Volatile Memory Express (NVMe)}}
NVMe is designed to overcome legacy protocol limitations by leveraging modern multi-core CPUs and abundant memory, establishing itself as the dominant standard for enterprise SSDs \cite{Sandisk}. Operating directly over PCIe, NVMe bypasses legacy I/O stacks designed for HDDs, enabling direct CPU-to-storage communication that eliminates controller overhead and minimizes latency \cite{FAST2018MQsim,ISCA2018FLIN,lerner2021not,sun2021convme}. Its architecture supports 64,000 independent command queues, enabling massive I/O parallelism and maximizing SSD utilization in multithreaded environments. By harnessing NAND flash's intrinsic parallelism, NVMe effectively eliminates traditional I/O bottlenecks.

\subsection{Advanced SSD Architecture}
Traditional SSDs operate as black boxes with internally managed operations, often leading to inefficient NAND utilization due to generic algorithms that cause unnecessary write amplification or underutilize flash endurance. This limitation stems from a lack of host-storage cooperation.
Next-generation architectures are redefining controller, NAND, and host interactions through designs such as Zoned Namespace (ZNS), Open-Channel SSDs (OCSSD), Flexible Data Placement (FDP), and computational storage drives. Software-defined SSDs further enable dynamic workload adaptation by abstracting flash management into programmable controllers with customizable firmware. These advances are blurring traditional boundaries between memory, storage, and computing.

\subsubsection{\textbf{Open-Channel SSDs (OCSSDs)}}
OCSSDs shift flash management tasks (wear leveling, garbage collection, address mapping) from the SSD controller to the host system. They expose raw NAND structures (channels, chips, dies, planes) to the host, enabling software (e.g. SPDK \cite{SPDK}) to control data placement and I/O scheduling for workload-specific optimization. This eliminates controller indirection, yielding deterministic latency and higher parallel I/O throughput. NVMe ZNS extends this model with sequential-write zones to reduce garbage collection overhead and improve endurance.
While OCSSDs enable superior performance and resource utilization, they introduce host-side complexity for low-level flash management. With improved standardization, OCSSDs bridge raw flash and application-specific storage solutions.

\subsubsection{\textbf{ZNS SSDs \cite{ATC2021ZNS,OSDI2023eZNS,OSDI2021ZNS+} }} ZNS SSDs organize storage into sequential append-only zones, directly addressing the inefficiencies of random writes in traditional SSDs. 
By enforcing sequential writes within zones, ZNS eliminates costly random-to-sequential translation, reducing latency and write amplification. Zones are erased as units, minimizing partial block erasures and enabling host-guided garbage collection for better wear leveling. However, this shifts the management complexity to the host, requiring careful zone allocation to avoid accelerated wear.
This zone-interface model represents a fundamental shift from traditional block-based SSDs, aligning software data placement with NAND flash's physical characteristics. 
Although block interfaces simplify host integration through LBA abstraction, they introduce significant overhead when handling random writes. ZNS instead exposes flash constraints to the host, enabling more efficient co-designed storage.

\subsubsection{\textbf{Flexible Data Placement (FDP) SSDs}} FDP SSDs offload logical-to-physical address mapping from the controller to the host, enabling granular data placement control. This reduces FTL overhead, especially benefiting write-intensive workloads. Unlike ZNS's sequential-write zones, FDP allows arbitrary data placement while maintaining host visibility into NAND structure. The host can implement intelligent wear-leveling strategies by distributing writes based on data access patterns.
Early implementations such as Samsung FDP SSDs\cite{SamsungFDP} show improved performance in cloud workloads, though they require host software support like Linux kernel drivers. With FDP now an approved NVMe standard and included in the Linux kernel, it is positioned for broader enterprise data center adoption where fine-grained storage control is critical.

\subsubsection{\textbf{Key-Value SSDs (KVSSDs)}}
KVSSDs have emerged as a critical component for distributed caching, LSM-tree databases, and real-time processing, as SSD architectures evolve toward application-specific customization. Traditional block-based SSDs perform poorly under key-value workloads due to frequent random updates, block abstraction mismatches, and inefficient FTL mapping, leading to high software stack overhead and severe write amplification.
KVSSD addresses these issues by natively storing data as key-value pairs and replacing traditional block drivers with a dedicated key-value interface. This design migates key-value operation primitives (i.e., $get()$, $put()$, and $delete()$) from the host to the SSD, enabling internal optimizations in FTL mapping, garbage collection, and write amplification control based on application patterns. The hardware-software co-design significantly reduces host CPU overhead. Samsung KVSSD \cite{SamsungKVSSD}, the industry's first commercial product in this category, implements a key-value interface over NVMe protocol.

In conclusion, SSD architecture design is evolving from closed, controller-centric models to disaggregated, host-aware systems that prioritize flexibility.
ZNS, Open Channel, and FDP architecture enable the host to align storage behavior with the applications \cite{chan2018hashkv,3736588,10.5555/2591272.2591299}.

\section{Challenges in SSD}\label{sec:challenges}
In this section, we give a bird’s-eye view of the challenges in SSDs, including reliability degradation, endurance limitations, latency volatility, and privacy and security challenges. 
These challenges create fertile ground for the innovation of device-side optimization technologies within SSDs.
\subsection{The Reliability Challenges}
The reliability of NAND flash memory is to accurately retrieve stored data after a period of time, despite program, erase or read operations or environmental changes during that time interval\cite{10693657,ye2024wearmeter}. 
Data storage in NAND relies on precise threshold voltages ($V_t$) control in memory cells~\cite{8734702}.
Bit errors occur when the read of a cell $V_t$ deviates from its target state, causing overlap between adjacent $V_t$ distributions. 
These shifts $V_t$, mainly due to program disturbance, read disturbance, and data retention issues, manifest in two main directions, where positive shifts towards higher threshold voltages and negative shifts toward lower threshold voltages. 
In the following, we will introduce the causes of different errors.

\textbf{Program errors} occur when memory cells fail to reach their target state during programming or when cell-to-cell interference arises from insufficient interleaving. 
Cell-to-cell interference is a phenomenon in which the threshold voltage of adjacent cells changes due to parasitic capacitance coupling during programming \cite{9580204}.
In incremental step-pulse programming, intermediate programming steps cause voltage threshold shifts in partially programmed cells, leading to misidentification of states and incorrect final programming from lower to higher voltage states (e.g., ER→P3 or P1→P2 in MLC) \cite{9473974}. These errors exhibit unidirectional upward voltage shifts and bypass ECC correction \cite{luo2016enabling}. 

\textbf{Erase errors} occur when an erase operation fails to reset a cell to the ER state when the applied voltage or duration is insufficient to fully remove trapped charges from the storage layer, leaving residual electrons that prevent the cell from returning to its intended threshold voltage level. 
This is due to the oxide degradation, charge trapping in the tunnel layer, or variations in process parameters that affect erase efficiency. 
Such failures may lead to read errors, reduced endurance, and increased program disturbance in subsequent operations, ultimately compromising the reliability of the memory device.

{\textbf{Data Retention Errors.}}
Retention errors result from $V_t$ reduction during idle periods due to charge leakage from tunnel oxide defects. Defects create conductive paths via Trap-Assisted Tunneling (TAT), enhancing low-field conductivity and generating Stress-Induced Leakage Current (SILC), which accelerates electron escape from the storage layer along the intrinsic electric field \cite{cai2017vulnerabilities}. Higher-voltage states exhibit the most significant $V_t$ shifts, leading to the highest error rates. Compared to planar NAND, 3D NAND shows accelerated early retention loss, with rapid $V_t$ shifts occurring within hours after programming due to fast de-trapping of charges near the storage layer surface \cite{LuoGCHM18,10.1145/3465332.3470877}.

{\textbf{Read Disturbance Errors.}}
Read Disturbance refers to abnormal $V_t$ shifts in victim cells during read operations, caused by unintended electron injection from the channel into the storage layer when a high pass-through voltage is applied to unselected wordlines \cite{8734702}. While a single read induces minimal shift, cumulative reads can cause significant errors \cite{8615714,8720454,KongZZCC20}. The phenomenon exhibits a dual effect: within a certain range, it compensates for data retention errors by offsetting electron loss through minor injection; beyond a threshold, however, excessive injection increases RBER \cite{10254389,9116337}. Cells near the ER/P1 distribution intersection are most sensitive due to process variations, with disturbance-prone cells clustering in the right tail of the ER state and resistant ones in the left tail of P1.
Overall, these errors exhibit linear growth with increasing P/E cycles, while the concurrent rise in interface state density at the channel-oxide interface further accelerates reliability degradation. Upon reaching a critical threshold, severe consequences emerge, including insufficient data retention below specifications, failures in program and erase operations, and uncorrectable read errors beyond ECC recovery capability \cite{9496603}.

\subsection{The Endurance Limitation}
As the number of P/E cycles increases in NAND flash memory cells, their reliability gradually deteriorates. 
Each P/E cycle degrades device performance due to electrical stress. 
Endurance is defined as the maximum P/E cycles a cell can withstand before failing to meet reliability metrics like data retention time and bit error rate.
When the actual P/E cycles exceed this rated value, the memory array may no longer guarantee reliable data storage and accurate read operations, and it is marked as a bad block by the SSD's internal FTL.
To replace these unusable blocks, SSDs typically employ over-provisioning (OP), allocating physical capacity exceeding their logically exported capacity. 
The number of bad blocks should surpass the over-provisioned spare capacity, then the SSD is deemed failed and reaches its end-of-life.
While early SLC flash stored a single bit per cell and lasted through thousands of P/E cycles, the push for greater capacity has led to MLC and TLC flash-based SSDs. 
Storing more bits per cell, however, comes at a steep cost to endurance, where each additional bit per cell reduces P/E cycle tolerance by approximately an order of magnitude~\cite{234980, li2022resolving,9006406}. 
Consequently, NAND flash endurance declines exponentially with higher cell density (SLC to PLC) and smaller process nodes, evidenced by sharply decreasing P/E cycles. 
The trend continues with QLC entering datacenters, and the planned PLC technology threatens to bring endurance limits down to mere tens or a few hundred cycles. 
Jeong et al. \cite{2591312} demonstrated that simultaneously extending erase duration and reducing erase or program voltage significantly enhances NAND flash endurance. The slower erase process using lower voltages effectively reduces damage to the tunnel oxide layer within NAND cells, since the magnitude of the erase voltage directly determines the erase depth of a flash block.

\subsection{The Latency Volatility}
In NAND flash memory, the latencies for read, program, and erase operations all increase with bits per cell, with erasing being the slowest, followed by programming, while reading remains the fastest \cite{7897088}.
Previous studies establish a clear latency hierarchy across NAND types~\cite{8739689,WeiLJZSL23}. SLC delivers the lowest latency (Program: 50–200 µs, Read: 20–25 µs, Erase: 2–5 ms). This escalates significantly in MLC (Program: 0.4–1.5 ms, Read: 40–110 µs, Erase: 5–10 ms) and further in TLC (Program: 0.8–2 ms, Read: 66–170 µs, Erase: 10–15 ms). QLC, at the density frontier, exhibits the highest latency (Program: 2–3 ms, Read: 120–200 µs, Erase: 15–20 ms).

The \textbf{program} latency $T_{prog}$  of NAND flash is directly proportional to the difference between the program start voltage ($V_{start}$) and program stop voltage ($V_{end}$), and also directly proportional to the number of incremental step pulse programming (ISPP) steps~\cite{2591312,10.1145/3599691.3603414}. 
Given constant $V_{start}$ and$V_{end}$, increasing ISPP steps extends programming time linearly. 
Conversely, with fixed ISPP steps, a larger voltage differential ($V_{end}$ - $V_{start}$ ) increases programming time proportionally \cite{2591312}. 
\textbf{Read latency} scales linearly with the number of reference voltages and retry operations. In TLC NAND, LSB page reads use one voltage for minimal latency, while MSB pages require four voltages, incurring significantly higher latency \cite{9251942}. Studies show SSDs at 3k P/E cycles require over three read retries on average, with each retry adding full-read equivalent latency, causing severe performance degradation \cite{3640372}.\textbf{ Erase latency} is dynamically affected by accumulated P/E cycles and process variation, requiring more iterative erase loops under ISPE as cells harden. Significant latency variations exist even among blocks at identical P/E cycles, showing 2.7 ms standard deviation in 3D TLC NAND \cite{10.1145/3620666.3651341}.
\textbf{ECC} significantly impacts SSD performance \cite{zhao2013ldpc,3357090,li2015maximizing}. Early in flash lifetime, low P/E cycles enable minimal-overhead hard-decision decoding. As reliability degrades with cycling, systems increasingly trigger high-latency soft-decision decoding with multi-voltage sensing and iterative computation. This shift is most severe at endurance limits, where read-intensive workloads suffer frequent soft-decision activations that degrade response time \cite{8101298}.

\subsection{Privacy and Security Challenges} 
The widespread adoption of SSDs introduces distinct security challenges that stem from their internal mechanisms. 
Persistent data retention until garbage collection complicates secure deletion. 
Encryption-based plausible deniability can be undermined by internal management operations. Additionally, SSD data handling methods may increase vulnerability to ransomware attacks while complicating data recovery efforts.

\textbf{Secure Deletion: }
SSDs widely store sensitive data but face architectural vulnerabilities enabling deleted data recovery. Encryption provides fundamental protection \cite{Reardon2012DataNE}, yet fails against coercive adversaries who can force key disclosure. Physical attacks on raw NAND chips can bypass encryption when hardware security is absent, particularly when both device and owner are compromised.
A critical limitation is SSD's inability to promptly sanitize data. Unlike HDDs supporting direct overwrites, SSDs employ out-of-place writes and wear-leveling that preserve obsolete data until garbage collection. This renders traditional overwrite-based sanitization ineffective. During updates, the FTL writes new data to different locations while original "digital remnants" persist in source blocks until erased, creating a data recovery window.

\textbf{Plausible Deniability Encryption: }
Encryption is essential for data privacy as it prevents unauthorized access by converting data into ciphertext. However, it is not impregnable. Under coercive threats where an adversary can obtain the key, the protection fails. Once the key is compromised, the encrypted data and its very existence are exposed, leading to potential theft.

In summary, the out-of-place update characteristic of SSDs and the complexity of the FTL make it impractical to directly apply traditional HDD-based  plausibly deniable encryption (PDE) \cite{chen2020infuse} schemes to SSDs. Achieving true plausible deniability encryption on SSDs requires delving into the low-level characteristics of flash media.

\textbf{Ransomware Defense: }
Ransomware remains a critical threat to data integrity \cite{park_RansomBlockerLowOverheadRansomwareProof_2019,kharraz_CuttingGordianKnot_2015}. Conventional software defenses like host-based intrusion detection systems are vulnerable to privilege escalation \cite{wang2006subvirt,continella2016shieldfs}, as advanced malware can obtain high-level privileges to disable protections and disrupt backups. This has driven interest in hardware-based protection integrated directly into SSDs \cite{tripathy_FormalModelingVerification_2023,yin_DefenseRecoveryStrategies_2025}.
The core opportunity for SSD-based ransomware defense lies in the "out-of-place update" mechanism of FTL \cite{gupta_DFTLFlashTranslation_2009}, which naturally preserves old data versions for recovery \cite{huang_FlashGuardLeveragingIntrinsic_2017}. However, significant challenges exist: SSD firmware operates at the LBA level without OS context (file names, process identities, or user intent) \cite{reidys_RSSDDefendRansomware_2022, ma_TravellingHypervisorSSD_2023}, creating a semantic gap that impedes differentiation between ransomware encryption and legitimate I/O-intensive tasks. Additionally, ransomware's random write patterns accelerate flash wear, and any retained-data defense mechanism may increase write amplification, reducing SSD lifetime.

\section{Device-Level Optimization Techniques}\label{sec:survey}
This section provides a comprehensive review of recent device-level optimization techniques developed to address these persistent challenges in flash-based storage systems. 
The relevant research landscape is analyzed through four key technical dimensions, which are advanced error correction mechanisms, innovations in the FTL design, advanced security schemes, and emerging system architectures, such as ZNS and FDP. 
For each category, we will categorize, compare, and discuss the state-of-the-art works, highlighting their underlying principles, advantages, and limitations.
\subsection{Error Correction Optimizations} 
\subsubsection{\textbf{Enhancing ECC Capability}}

Low-density parity-check (LDPC) codes are the most commonly used ECC in NAND flash memory storage. To enhance the read response speed and reliability of SSDs, extensive research has focused on optimizing LDPC read performance. These strategies primarily fall into three categories: optimizing read voltage strategies, improving decoding algorithms, and innovating storage architectures or data management.

\textbf{Voltage Optimization Schemes:} Zhang et al. \cite{8478389} developed a dynamic sensing voltage adjustment scheme using RBER variations with retention time/P/E cycles. Zhao et al. \cite{zhao2013ldpc} introduced fine-grained LDPC progressive reading, applying finer voltages only after read failures. Bao et al. \cite{10318173} improved this with progressive adaptive granularity decoding to optimize response time. Du et al. \cite{10032056,9933162} proposed predictive fine-grained reading for 3D NAND layer variations, later extending to latency-aware multi-granularity LDPC reading \cite{10.1145/3585075}.

\textbf{LDPC Decoding Enhancements: }Du et al. \cite{8509142} utilized MLC pair-bit errors for PBE-aware decoding acceleration. Wu et al. \cite{8634908} optimized LLR tuning for 3D TLC error patterns. Zhang et al. \cite{8119288} accelerated MSB decoding using LSB page results, while Li et al. \cite{7858383} adjusted read voltage via abnormal decoding behavior modeling. Zhang et al. \cite{10.1093/comjnl/bxad069} combined duplication coding with LDPC soft-decision decoding.

\textbf{Data Organization Strategies: }Li et al. \cite{li2019leveraging} left approximate data unprotected to enhance error resilience. Di et al. \cite{10.1145/3194554.3194571} employed shortened data schemes with low-cost ECC. Zhao et al. \cite{6855550} used data interleaving and dependency-aware decoding. Nie et al. \cite{Nie2024} grouped high-locality pages across RBER-varied physical pages to reduce read retries. Du et al. \cite{3317759} implemented multi-granularity LDPC for 3D layer variations. Lv et al. \cite{8834728} designed smart refresh for tail latency reduction, and Li et al. \cite{li2018selective} selectively compressed high-read-frequency pages.

\subsubsection{\textbf{Accelerating ECC Process}}
ECC decoding optimization employs dynamic prechecking and adaptive engines to streamline decoding processes. Content-aware correction tailors programming parameters and ECC strength to data characteristics. For error-tolerant data, approximate writes accelerate operations while maintaining reliability. These methods can reduce latency while preserving data integrity in NAND flash systems.

\textbf{ECC Decoding Optimization: } Du et al. \cite{10.1145/3585075} employ dynamic error prechecking with sub-page partitioning to bypass redundant decoding for error-free data. DEPS \cite{9091550} utilizes multi-granularity LDPC engines that adapt error correction strength to layer-specific characteristics. For MSB pages with higher error rates, intelligent voltage adjustment \cite{10371374} reduces hard decoding iterations.

\textbf{Content-Aware Error Correction: } CDBER \cite{6974682} exploits content-dependent bit error rate variations by dynamically adjusting programming speeds for lower-error pages. 

\subsection{Flash Translation Layer Enhancements}
\subsubsection{\textbf{Data Reliability Mechanisms}}
In this section, we will review reliability enhancement technologies at the SSD device level, beyond ECC optimization capabilities. First, we will explore fine‑tuned voltage applying methods and intelligent data placement strategies to improve data reliability at the stage of programming. Next, for data stored over time, we will discuss refined read voltage adjustment strategies and analyze proactive maintenance approaches based on refresh mechanisms and internal RAID techniques. Finally, we will survey failure prediction technologies, during the later stages of device service lifetime.

\textbf{Operating Voltage Optimization Techniques: }
Several studies have demonstrated that reducing the high pass-through voltage ($V_{pass}$) within memory blocks can effectively minimize the unintended impact of read operations on adjacent unread cells. 
Ha et al. \cite{MitigateRDError} narrow threshold voltage distributions during programming to inherently lower $V_{pass}$ requirements. Li et al. \cite{8064482} selectively reduces $V_{pass}$ in specific wordlines based on read-hotness awareness , while Wu et al. \cite{10777429} improve read voltage calibration under data scarcity using P/E cycle-augmented ML models.
Zhang et al. mathematically fit voltage distributions for direct read voltage calculation \cite{9241228}, while novel mechanisms like GIDL current modulation suppress hot carrier injection \cite{10138658} and delayed rereading mitigates temporary errors \cite{8956089}.
Kong et al. observe that the RBER caused by data retention issues can improve following certain read operations, regardless of the block's prior P/E cycle history~\cite{KongZZCC20}. 

\textbf{Enhancing Reliability via Data Allocation Strategies: }
Within an SSD, individual blocks exhibit varying levels of reliability, and the overall reliability of the SSD is often limited by its least reliable blocks, where the limitation becomes more significant as the SSD aged. 
To address this "weakest‑link" effect, researchers have developed data allocation strategies that take into account both data access patterns and the characteristics of individual block, thus preventing less-reliable blocks from failing prematurely.
Zhao et al.~\cite{BlockawareRD} proposed a block-aware strategy that preferentially stores hot data in blocks with lower historical read counts while directing less frequently accessed data to blocks that have endured more read operations.
Liao et al. \cite{RERF} used predicted future read counts to assign hot data to high-endurance blocks. 
Jun et al. \cite{li2020mitigating} combined both read count and P/E cycle history for data placement decisions.
DRR \cite{10.1145/3386263.3406921} reduces relocation overhead in 3D NAND by routing reads via parity redundancy and data duplication to less-disturbed blocks, using real-time I/O patterns to minimize data migration.
Cocktail \cite{9919274} employs mixed cold-hot data allocation, pre-filling blocks with cold data while dynamically allocating space for hot data to prevent concentration effects and reduce refresh frequency.

\textbf{Optimizing Refresh Operations: }
 Refresh operation serves as a preventive data maintenance mechanism in SSDs, designed to proactively preserve data integrity. It works by silently migrating and reprogramming data in the background before uncorrectable errors occur. However, this process contributes to write amplification, accelerates flash wear, and degrades the performance. Consequently, research efforts are largely directed toward optimized trigger strategies to minimize the overall effect on SSD endurance and performance.
Li et al. \cite{8119291} group blocks by reliability level and placing hot data in reliable blocks, optimizing read performance through heterogeneity-aware data placement.
Cui et al. \cite{10.1145/3386263.3406921} replace fixed refresh cycles with error rate-based dynamic adjustment, monitoring the state of the block and data hotness to avoid unnecessary refreshes.
With the continuous increase in flash memory density, the data retention time has been significantly shortened, while traditional periodic refresh techniques lead to notable performance degradation and write amplification.
ApproxRefresh \cite{10.1145/3372799.3394362} leverages error tolerance in approximate computing to reuse uncorrectable data, implementing mapping-free refresh with ECC bit updates instead of data movement. 
Cui et al. \cite{9496603} similarly focus on mitigating write amplification and performance loss caused by periodic refresh operations, particularly considering that some uncorrectable data can be tolerated by certain applications.

\textbf{The Redundant Array of Independent Disks (RAID) Techniques: }
To address the reliability requirements of enterprise storage, the application of traditional RAID technology to SSDs has become a standard practice for building highly reliable storage systems. However, this integration is far from a simple technological migration; it requires a careful balance among performance, security, cost, and storage efficiency. This section will explore the application of RAID technology in SSD environments.
Common configurations in RAID technology is including RAID 0 \cite{50214} for performance via stripping without redundancy; RAID 1 \cite{298619} for integrity through capacity cost mirroring; RAID 5/6 \cite{1807061,inproceedings} balances redundancy and efficiency via distributed parity for QLC or TLC arrays.
RAID improves SSD reliability through multi-drive redundancy and checksum-based integrity protection, while parallel architectures leverage NVMe concurrency for bandwidth gains. However, checksum-induced read-modify-write cycles degrade write efficiency \cite{3640368}.
Asymmetric-RAID \cite{3665952} for heterogeneous SSDs via asymmetric striping and performance-aware address space allocation.
The hybrid RAID \cite{10604932} intelligently routes full-stripe writes to RAID5 and partial/small writes to RAID1. 
Li et al. \cite{3596274} reduce WAF via replication buffering, while SSRAID \cite{10636805} resolves thread contention via stripe-queued architecture.

Beyond multi-SSD RAID configurations, internal components like memory chips and planes can be organized in RAID-like topologies to enhance reliability and performance.
Kim et al. \cite{10.1145/2349896.2349900} proposed DYS-RAID with dynamic variable-size striping to reduce small-write overhead. 
Im et al. \cite{5496997} designed a flash-optimized RAID-5 using delayed parity updates and partial caching to minimize write amplification.
PTan et al. \cite{7255200} created H2-SSD using heterogeneous SLC or MLC chips and asymmetric parity to optimize endurance. 
Lee et al. \cite{6575359} analyzed the ECC-RAID tradeoffs and implemented eSAP for adaptive redundancy. 
Im et al. \cite{9211551} designed patch-based buffer management with selective parity cache and adaptive scheduling to reduce P/E operations.

\textbf{The SSD Failures Prediction Techniques: } 
SSD failure prediction technology aims to provide a critical early warning window for data backup, service migration, and hardware replacement before irreversible failure occurs. This is especially crucial in data centers with large volumes of SSDs, to prevent fail-stop failures from causing server downtime and to help optimize resource provisioning.
Field data reveal that real-world annualized failure rates (AFRs) often deviate from manufacturer specifications, highlighting how workload intensity and operational environments accelerate failure modes\cite{1267905,1288785}.
Lu et al. \cite{280680} studied fail-stop and fail-slow patterns in NVMe SSDs, finding distinct behaviors compared to SATA SSDs, including infant mortality differences, WAF-failure rate relationships, and high correlation among co-located drive failures. 
Alter et al. \cite{10902250} developed ML models to analyze workload-triggered failures and distinguish infant mortality from later-life failures. 
Chakraborttii et al. \cite{3421300} proposed 1-class ML methods trained only on healthy data to handle imbalanced datasets, introducing a 1-class autoencoder for robust prediction. 
Zhou et al. \cite{9521302} used VAE-LSTM for unsupervised failure detection by learning normal behavior patterns.
Zhang et al. \cite{285760} proposed MVTRF, a multiview multitask random forest integrating histogram and sequence features to predict failures, classify manifestations, and estimate lifetime. 
Ha et al. \cite{298619} designed RLW, a reinforcement learning-based watchdog for real-time failure detection via VFS integration. 
Jiang et al. \cite{10467231} applied model-wise class balancing to address dataset imbalance.

\subsubsection{\textbf{Endurance Enhancement Techniques}}
This section will outline the techniques to enhance the endurance of the SSD by optimizing the programming code methods and leveraging intrinsic physical properties through self-healing techniques, then it further focuses on minimizing write amplification through effective garbage collection and wear-leveling algorithms. 
These methods aim to enhance the SSD's endurance to withstand a substantially higher volume of data writes.

\textbf{Coding Optimization:}
Advanced coding schemes directly combat the endurance limitation of the flash memory by restructuring how data is written and managed at the bit and page levels. 
Traditional Write-Once Memory (WOM) codes are inefficient in multi-level flash memory, which can improve logical data write volume by at most 50\%, failing to meet the demands of high-density SSDs \cite{188452}. WOM-v breaks through traditional bit-level constraints by directly utilizing the 16 voltage levels of QLC cells as encoding states, allowing voltage levels to increase monotonically while supporting more flexible state transitions \cite{254260}.
The WOM-FTL scheme integrates the WOM-v code with the FTL in the high-density flash memory, by leveraging the partial in-place update capability of WOM-v codes to achieve efficient secure deletion \cite{11007599}.
Fixed Gray code encoding fails to balance performance and lifetime, as frequent rewrite operations exacerbate flash wear and shorten SSD endurance. MGC dynamically selects highly reliable Gray codes at different stages of SSD usage to reduce rewrite operations \cite{10070946}.
C. Gao et al. proposed a novel reprogramming scheme, which allows multiple programming operations on a cell before erasure by initially configuring TLC cells in MLC mode and achieving two additional reprogramming cycles through voltage state migration, enabling up to three programming cycles per cell \cite{gao2019constructing,gao2022reprogramming}.

\textbf{Self-healing Techniques: }
Several research efforts have investigated self-healing techniques to enhance flash memory endurance by facilitating the removal of trapped electrons in the oxide layer, thereby improving reliability.
Research on self-healing techniques focuses on removing trapped electrons to enhance flash endurance, primarily through thermal-assisted and passive healing methods. Thermal acceleration uses controlled heating to speed up electron dissipation, with implementations including batch-based heating \cite{chang2017relieving} and adaptive algorithms that consider P/E cycles and workloads \cite{cui2020smartheating}. However, this approach requires valid data migration before heating and may increase error rates in nearby data.
Alternative passive methods optimize dwell time between erase operations to naturally promote electron recovery. Studies \cite{wu2011exploiting} confirm longer dwell periods improve healing, while Luo et al. \cite{luo2018heatwatch} developed adaptive read voltage adjustments based on measured healing intervals. Mohan et al. \cite{mohan2010learned} further modeled the relationship between stress conditions and self-healing efficacy.

\textbf{Enhanced Garbage Collection and Wear Leveling: }
Recent research has significantly improved SSD endurance through optimized garbage collection and wear leveling techniques, with innovations including the time harmonization strategy \cite{9076275} that implements lifespan-aware data management, BER \cite{9104673} that enhances durability through system-level wear leveling, GuardedErase \cite{277832} that employs dual-mode erasure to protect vulnerable wordlines, and AERO \cite{10.1145/3620666.3651341} dynamically adjusts erase latency. The SWAN architecture \cite{10.5555/3358807.3358875} isolates I/O and garbage collection operations, while LLSM \cite{9852747} provides lifespan optimization for LSM-tree systems through tier-aware allocation. Jiao et al. \cite{10.1145/3538643.3539750} propose a capacity-variance design to avoid write amplification, and Ada-WL \cite{10527393} adopts LSTM-based wear-leveling for uneven workload distributions.
STRAW's wordline-level disturbance modeling and dynamic migration overcoming the conservatism of traditional block-level management and significantly reducing unnecessary migration operations \cite{10795178}.
To address the endurance imbalance across layers in 3D NAND flash, the LA-Write strategy directs write operations to more durable layers via a write-skip unit and a layer-aware probability table, effectively balancing wear \cite{10781366}.
This study proposes an application/SSD co-designed read latency relaxation strategy that uses performance hints to selectively rewrite only high-latency data, reducing unnecessary background operations \cite{10.1145/3445814.3446733}.

\textbf{Proactive Bad Block Management: }
Traditional bad block management employs a static replacement mechanism, which is overly conservative and does not fully exploit the endurance potential of SSDs. Current research explores finer-grained bad block management strategies, adopting a trading capacity for endurance approach that allows continued use of the remaining healthy space.
WAS addresses premature failure of weak blocks caused by varying wear tolerance by implementing a weak-page RBER-based wear detection mechanism, thus effectively isolating weak blocks and maximizing the endurance potential of strong blocks\cite{8806851}.
Yen et al. discovered spatial correlation in error behaviors among physically adjacent flash memory blocks, and proposed that when any bad block is detected within a cluster, the entire adjacent block cluster should be collectively retired to preemptively mitigate potential failure risks\cite{9773252}.
CVSS dynamically reduces the logical capacity, allowing the SSD to mask high-error-rate blocks, thereby trading the capacity for extended lifetime\cite{294793}.
LaVA proposes a layer-aware bad block management technique that employs layer-granular management\cite{10546546}.
Salamande divides the SSD into multiple logical mini-disks to align with hardware fault granularity. Additionally, it reallocates the data bits of failed pages as extra ECC bits, further leveraging worn flash memory\cite{10.1145/3713082.3730386}.

\subsubsection{\textbf{Performance Improvement Techniques}}
For latency reduction, simplified read-retry with predictive calibration minimizes read disturbance recovery time by proactively adjusting read thresholds. To improve I/O concurrency, parallel architectures leverage multi-channel striping and interleaved data placement, enabling high throughput under concurrent workloads. Additionally, tiered caching mechanisms mitigate slow-access latency by dynamically retaining hot data in faster storage tiers. Adaptive garbage collection and wear-leveling algorithms effectively mitigate the latency induced by intensive garbage collection operations and frequent page migrations.

\textbf{Simplified Read-retry Process: }
Research focuses on minimizing read latency by streamlining retry processes. CACHE READ \cite{10.1145/3445814.3446719} improves retry efficiency through pipelined execution that overlaps sensing and data transfer phases, combined with adaptive sensing time reduction. For near-zero retry objectives, ORVD-WRRO \cite{LI2022114509} eliminates traditional retry mechanisms in 3D NAND by embedding voltage calibration within standard reads using Overlap Error Recording ECC, enabling single-cycle optimization. Ye et al. \cite{3640372} implement predictive voltage calibration via LDPC syndrome-guided retry tables, achieving near-optimal voltage selection that virtually eliminates retries. This evolution from accelerating retries to preventing their occurrence represents a paradigm shift in read optimization methodologies.

\textbf{Optimized Parallelism and Scheduling: }
An SSD typically comprises multiple layers of hardware parallel units, such as flash channels, chips, dies, and planes. Current research focuses on enhancing the parallel processing capability of SSDs, enabling multiple parallel units to perform read, program and erase operations simultaneously. This significantly improves overall bandwidth, achieving near-linear performance scaling. 
To address parallelism limitations in high-density 3D NAND, SOML read \cite{10.1145/3297858.3304035} enables simultaneous subpage retrieval from single wordlines, while intelligent request merging reduces redundant operations. Parallel Read Retry \cite{9829889} implements simultaneous retry processing with adaptive voltage calibration. For compression bottlenecks, DPC \cite{9783102} with split FTL enables fine-grained data distribution across channels. Do et al. \cite{10.1007/s00778-020-00648-z} offload LSM-tree management to SSD controllers via batched I/O, eliminating host-storage redundancy.
DLV scheduler \cite{8081787} prioritizes low-latency accesses while accounting for process variation and aging effects. Lv et al. \cite{9218540} tackle read-write interference through spatial partitioning and dynamic migration. For QLC latency variations, RRTS \cite{10979979} dynamically selects low-latency page combinations and optimizes parity placement. Li et al. \cite{8362696} introduce dynamic cost adjustment based on access patterns, applying higher write costs for read-intensive data to optimize read speed.

Modern storage systems, especially with high-density 3D NAND technologies, face significant I/O scheduling challenges due to latency variations, interference effects, and aging-related degradation, leading to the development of advanced techniques that intelligently exploit latency heterogeneity and optimize resource allocation.
Recognizing the dramatic latency differences in 3D NAND cells, some systems implement Retention-Latency Variation (RLV) management by identifying hot data and relocating it to faster pages while employing cache optimizations to minimize slow-page accesses \cite{10.1145/3386263.3406953}. ReadGuard \cite{10.1145/3676884} implements block-level latency profiling for priority-aware scheduling. PA-SSD \cite{8798696} coordinates homogeneous page-type allocation and prioritizes faster page accesses. Wu et al. \cite{8823009} propose a reinforcement learning framework that dynamically adjusts I/O merging policies using Q-learning to balance throughput and latency variance.
To address the issue of multi-layer I/O interference caused by background tasks in storage systems, HuFu architectureoffloads background I/O scheduling to the SSD, leveraging the SSD's global view and flash resource control capabilities to achieve coordination between prioritized foreground I/O processing and opportunistic background I/O scheduling \cite{10159481}.

\textbf{Caching Strategies for Latency Variation Mitigation: }
The inherent latency variations in modern flash memory architectures, particularly in QLC-based storage systems where read delays can vary significantly across different page types, have prompted innovative caching strategies to mitigate performance bottlenecks.
PACA \cite{9978511} employs dynamic cache partitioning to selectively cache high-latency page data with lifecycle-aware metadata management, while Shi et al. \cite{shi2022read} propose an integrated data placement framework that allocates hot data to fast pages, relocates cold data to slower regions, and maintains an auxiliary cache for slow-page data. DAC \cite{SUN2023102896} dynamically adjusts cache allocation between hot/cold data and proactively generates clean pages via active write-back to boost replacement efficiency, and FastCache \cite{QIAN2022102718} addresses concurrent microwrites through a two-level cache structure with optimistic locking and variable-position merging to reduce fragmentation and improve throughput.

\textbf{Optimized Garbage Collection and Wear Leveling: }
In conventional SSD architectures, garbage collection, wear leveling, and I/O requests compete for shared system bus and DRAM resources. Data migration, which is inherently required for garbage collection and wear leveling, involves moving data between NAND flash and off-chip DRAM. This process consequently degrades overall performance and increases tail latency.
Recent research has advanced garbage collection and wear leveling techniques through victim block selection algorithms that minimize the overhead of valid page migrations. 
First, victim block selection algorithms aim to identify optimal candidates for erasure. 
Lin et al. \cite{lin2015dynamic} used historical data update patterns to enhance both garbage collection efficiency and fairness.
Yang et al. \cite{yang2019reducing} reduced overhead by predicting data hotness using LSTM networks to forecast future access patterns across temporal and spatial dimensions.
Second, data grouping strategies organize information with similar characteristics into distinct streams. 
Multi-stream SSD \cite{kang2014multi} and lifetime-based grouping \cite{wang2022separating} segregate data by characteristics.
Ren et al. \cite{10745809} proposed a victim block selection strategy that jointly considers the block dwell time and invalid page counts.
Du et al. \cite{mi12070846} addressed SSD performance cliffs by distributing page migrations via PreGC, reducing tail latency with minimal write amplification.
To address resource contention, rCPB \cite{8863524} enhanced copy-back operations and Decoupled SSD \cite{10.1145/3579371.3589096} employed network-on-chip solutions to reduce I/O interference. 
For wear leveling innovations, the DWR scheme \cite{9774738} addresses the significant increase in read latency by partitioning the flash memory area into a high-wear area for write operations and a low-wear area dedicated to storing hot read data. 
Further research design adaptive physical partitioning with dynamic capacity adjustment and least-recently-used (LRU) based data migration to balance endurance and speed \cite{song2023adaptive}.

\subsection{Advanced Emerging Architectures}

\subsubsection{\textbf{Techniques for Endurance Enhancements}}
Emerging SSD architectures enable hardware-software co-design to overcome endurance limitations. This section reviews key techniques that leverage these advancements to enhance device endurance.

\textbf{ZNS SSD: }
To address write amplification, which is critical for impact on endurance in ZNS SSDs, a space-aware garbage collection strategy \cite{HotStorage2023FAR} dynamically adapts zone reset frequency based on available free space. Garbage collection operations are reduced under high space availability to minimize unnecessary data migration, while being increased when space is constrained to ensure capacity. This approach significantly extends SSD endurance while maintaining performance.
ZNS SSDs also face bottlenecks in RAID and compression scenarios, such as frequent parity updates and compression index overhead.
Zebra \cite{HPCA2025Zebra} uses the ZRWA feature to achieve efficient parity updates and avoid the write amplification of traditional RAID.
CCZNS \cite{HPCA2025CCZNS} separates the compression and indexing processes, and the host and SSD jointly manage compressed data to improve space utilization.
These approaches help address the unique challenges of managing data in zoned storage environments.

\textbf{Key-Value SSDs: }
Using a native key-value interface to minimize redundant writes, KV-SDs effectively reduce garbage collection overhead, a primary factor that affects SSD's endurance.
Several studies optimize KV-SSDs by aligning data management with device characteristics. Oh et al. \cite{oh2021efficient} organize data by key-value traits to improve space efficiency and reduce garbage collection overhead, while Wu et al. \cite{wu2020integrating} use FTL mapping for copy-free compaction and direct flash access, cutting write amplification. As LSM-trees are HDD-originated, PLSC-tree \cite{chen2020parallel} introduces a two-layer KV-SSD management strategy to boost writes via parallel logging and reduce garbage collection impact. KVRAID \cite{qin2021kvraid} groups similarly-sized objects to limit object amplification.
For ZNS SSDs, LSM-trees’ sequential writes align well with zoned constraints. Research focuses on garbage collection and compaction enhancements: Lu et al. \cite{lu2022revisiting} propose zone-aware garbage collection, and LifetimeKV introduces lifetime-aware compaction to balance SSTable aging across levels.

\subsubsection{\textbf{Techniques for Performance Improvement}}
The emergence of advanced SSD architectures introduces new opportunities for performance optimization by enabling closer hardware-software co-design. This section reviews key techniques that leverage these architectures to address specific performance bottlenecks. 

\textbf{ZNS SSDs: }
ZNS SSDs face performance challenges, including I/O unfairness (e.g., long writes blocking short requests) and fragmentation. Fair-ZNS \cite{TCAD2024Fair-ZNS} introduces a self-balancing scheduler to prioritize delayed requests while maintaining throughput. eZNS \cite{OSDI2023eZNS} and FlexZNS \cite{FlexZNS} use dynamic zone management (e.g., elastic sizing) to reduce fragmentation. To improve garbage collection, Brick-ZNS \cite{TACO2024Brick-ZNS} migrates data internally via ZNS commands, avoiding host-device transfers, while Seo et al. \cite{HotStorage2023gcZNS} enhance F2FS segment allocation with parallel garbage collection for lower latency.

\textbf{Key-Value SSDs: }
KVSSDs introduce challenges in garbage collection, write amplification, and tail latency. PinK \cite{ATC2020PinK} adopts an LSM-tree design to improve tail latency in hash-based KVSSDs, while Vigil-KV \cite{ATC2022Vigil-KV} employs software-hardware co-design for further read tail latency optimization. These approaches demonstrate the need for specialized solutions beyond traditional SSD-based key-value systems.
Parallelism optimization has been a key focus for KV storage performance. SplitZNS \cite{huang2023splitzns} creates smaller zones aligned with LSM-tree levels to exploit SSD parallelism. Wang et al. \cite{wang2014efficient} optimize I/O scheduling for open-channel SSDs to maximize throughput, while Lee et al. \cite{lee2023waltz} use ZNS's zone append command to reduce LSM-tree compaction tail latency through parallel writes.

\textbf{Hybrid and Convertible SSDs: }
Hybrid SSD research focuses on using fast SLC/MLC modes for caching while employing QLC for bulk storage \cite{AlsalibiMAS18}. A key challenge is capacity balancing, requiring choices between static partitioning with fixed ratios \cite{AlsalibiMAS18} or dynamic reconfiguration that adapts to usage patterns \cite{LuoLLS23,WangCNLY24,Intel}.
A second challenge involves mitigating throughput degradation during data migration from fast to dense tiers. Solutions include tiered hierarchies using MLC as intermediate buffers to distribute migration overhead, and I/O profiling that directs high-velocity writes directly to QLC to bypass caching layers \cite{LuoLLS23}.
Third, garbage collection optimization in heterogeneous memory is challenging as conventional algorithms struggle with uneven write distribution in SLC regions. Solutions include runtime adaptation that temporarily converts QLC blocks to SLC mode during low-efficiency periods \cite{ShiLLLLS21}, and reinforcement learning approaches for intelligent collection target selection \cite{WeiLJZSL23}.

\textbf{CXL SSDs: }
SrNAND \cite{11006506} enhances CXL SSD small-data read bandwidth through partial reads with two-stage ECC, merged sense/data operations for parallel transfer, and request merging to avoid page redundancies. ExPAND \cite{kwon2023cache} offloads prefetching to CXL-SSDs using ML-based prediction and topology-aware scheduling, improving performance for graph/SPEC workloads.

\subsection{Solutions to Privacy and Security Issues} 
Modern SSD architectures introduce unique security vulnerabilities that require device-level solutions. This section surveys defenses against these threats. It examines secure deletion techniques for flash media, PDE systems for SSDs, and ransomware defenses leveraging intrinsic SSD operations. Collectively, these approaches represent a shift towards embedding robust security primitives directly into the storage controller.

\subsubsection{\textbf{Secure Deletion in SSD}}
Researchers have developed device-level optimizations to address SSD-specific security challenges. These solutions leverage intrinsic flash memory properties rather than adapting HDD-era methods, and fall into three categories, which are encryption-based, scrubbing-based approaches and hybrid frameworks of sanitization.

\textbf{Scrubbing-based Sanitization Schemes:} Since software-based overwriting is ineffective, reliable data erasure on SSDs requires FTL-aware mechanisms. A core technique is scrubbing~\cite{gao2019constructing}, which sanitizes data via in-place page reprogramming, avoiding full block erasure. However, it induces program disturbance in adjacent cells, necessitating preemptive page migration that incurs write amplification and performance overhead.
Multiple device-level techniques have been proposed for secure SSD erasure. Early work includes an ISPP scheme for precise voltage control \cite{suh19953} and an interference-aware method for 3D NAND \cite{wang2018scrubbing}. Recent innovations leverage coding theory, such as FSD using WOM-v codes for in-place overwrites \cite{cui2023fast}, and machine learning, exemplified by PollSan's dynamic erasure framework \cite{wu2024polling}.

\textbf{Encryption-based Sanitization Schemes:}
To circumvent the performance and physical limitations of physical erasure, encryption-based sanitization (crypto-shredding) has emerged as a key alternative for SSDs. It encrypts data before NAND writing, achieving secure deletion by destroying the small cryptographic key, rendering the ciphertext permanently unreadable \cite{braga2014adding}. This concept has been refined with granular key management \cite{Reardon2012DataNE, wang2018novel} and developer libraries \cite{yang2018sadus}. To close the vulnerability window of delayed key deletion, FlashFox \cite{cheng2025flashfox} actively corrupts data via secret sharing for immediate destruction. For ZNS SSDs with large zone sizes, PRESS \cite{hsieh2024press} uses persistence relaxation and host DRAM buffering to enable efficient, near-zero-cost key deletion.

\textbf{Hybrid Frameworks of Sanitization:} 
To overcome the trade-offs of single sanitization methods, hybrid frameworks have been developed.
ErasuCrypto \cite{liu2017erasucrypto} dynamically selects the most efficient method between cryptographic and physical erasure, while SDDK \cite{xiong2020secure} employs a hybrid key-destruction and partial-erasure strategy for IoT devices. 
TedFlash \cite{zhang2018ensuring} uses random data placement to ensure the final storage state reveals no operational history.

\subsubsection{\textbf{Plausibly Deniable Encryption inside SSD}}
To defend against powerful adversaries, PDE protects against powerful adversaries by concealing its very existence of sensitive data. It enables a ciphertext to be decrypted into either the true sensitive data with the real key or a credible decoy with a decoy key. If coerced, the user can disclose the decoy key, thereby denying the existence of any hidden data without revealing the true information.

\textbf{Adversarial model:}
PDE typically relies on several common adversarial assumptions: the attacker gains physical SSD access via standard interface commands after the device is unmounted or powered off, bypassing file system and block layers, but cannot capture runtime states; the adversary coerces key disclosure but ceases if no sensitive data evidence is found; and while aware of PDE design existence, the adversary only escalates coercion upon confirming actual sensitive data beyond initially provided decoy keys.

\textbf{Filesystem-level PDE:}
Filesystem integration systems integrate PDE mechanisms directly into the YAFFS2, e.g., DEFY~\cite{Peters2015}, INFUSE~\cite{chen2020infuse}. They provide structured hidden volumes and deniable access within the filesystem layer. While conceptually sound, their practical adoption is severely limited because YAFFS2 is primarily used in niche embedded flash systems and is rarely found in modern consumer or enterprise SSDs.

\textbf{Flash cell-level PDE:}
Early PDE schemes exploited physical flash properties. Wang et al. \cite{wang2013hiding} leveraged programming-time variations for covert channels, though slowly and with high wear. VT-HI \cite{Zuck2018} encoded data in SLC voltage levels—efficient but flash-specific. Hide-and-Seek \cite{raquibuzzaman2023hide} modified threshold voltages, yet incurred latency and limited capacity by embedding only in "0" bits of public data.

\textbf{FTL-level PDE:}
FTL-centric PDE schemes leverage the flash translation layer for deniability. DEFTL \cite{Jia2017} first integrated a hidden volume into the FTL to resist single-snapshot adversaries, though it remained vulnerable to multi-snapshot analysis. MDEFTL \cite{jia2021mdeftl} enhanced security against such threats using dummy writes and randomized allocation, but incurred significant write overhead and P/E cycle consumption.

\subsubsection{\textbf{Ransomware Defense}}
To counter the challenges posed by ransomware, a rich body of research has emerged focusing on firmware-level defense and recovery techniques within SSDs. These approaches leverage the intrinsic properties of NAND flash to create a secure, hardware-isolated last line of defense. The evolution of these techniques can be broadly categorized into several key strategies.

\textbf{Foundational Passive Retention and Recovery:}
Early strategies used the out-of-place updates for data recovery. FlashGuard ~\cite{huang_FlashGuardLeveragingIntrinsic_2017} pioneered this by modifying garbage collection to save any page that was read and then updated, assuming ransomware reads data before encrypting it. While this ensures perfect recovery, it causes high storage and performance overhead. Expanding on this, TimeSSD ~\cite{wang_ProjectAlmanacTimeTraveling_2019} provides a broader time-travel capability, retaining all data versions for a set period using back-pointers and delta compression to manage the overhead.

\textbf{In-SSD I/O Pattern-Based Detection:} 
To reduce overhead, later research integrated active detection into the FTL. SSD-Insider~\cite{baek_SSDInsiderInternalDefense_2018} pioneered this by using a decision tree to identify ransomware based on I/O patterns, enabling more selective data retention. MimosaFTL \cite{wang_MimosaFTLAddingSecure_2019} refined this with K-means clustering for better pattern detection and added a precise binary-search recovery method. To counter more advanced threats, SSD-Insider++ \cite{baek_SSDAssistedRansomwareDetection_2021} improved the model by treating TRIM commands as part of a potential attack and added a "lazy detection" feature that compares the entropy of old and new data before erasure, providing an extra security layer against evasive ransomware.

\textbf{In-SSD Content-Aware Detection:} 
To overcome the limits of pattern-based detection, some solutions inspect data content directly. RansomBlocker~\cite{park_RansomBlockerLowOverheadRansomwareProof_2019} uses a two-phase approach: a fast entropy filter first identifies suspicious data, which a more accurate CNN classifier then analyzes to minimize false positives. AMOEBA~\cite{min_ContentBasedRansomwareDetection_2022} introduced a custom DMA hardware engine to perform high-speed, inline entropy and similarity checks during data transfer. AMOEBA combines these content metrics with I/O intensity into a single risk score for highly accurate and efficient detection.

\textbf{Advanced Co-Design Architectures:}
Recognizing the inherent limits of in-SSD solutions, the most recent works propose advanced architectures that coordinate defense across system layers. RSSD~\cite{reidys_RSSDDefendRansomware_2022} introduces a network-storage co-design that securely offloads data versions and logs to remote cloud storage, enabling more powerful offline attack analysis. RANSOMTAG~\cite{ma_TravellingHypervisorSSD_2023} proposes a Hypervisor-SSD co-design to perform fine-grained file-level recovery, moving the detection logic to a hypervisor to leverage the host CPU power and OS-level context.

\section{Open Research Challenges}\label{sec:open}
\subsection{\textbf{Scalability Issues with QLC/PLC NAND}} 
The transition from TLC to QLC and PLC NAND has significantly increased storage density, but introduces critical scalability challenges that impact reliability, performance, and cost efficiency. This subsection examines these challenges and highlights key research directions for future SSD optimizations from device level.  

\textbf{Endurance and Write Performance Degradation:} QLC/PLC NAND stores 4-5 bits per cell, drastically reducing write endurance (typically 100-1,000 P/E cycles for PLC compared to 3,000-10,000 for TLC). The increased program interference and write amplification further accelerate wear-out. Future research could explore adaptive wear-leveling schemes that account for ultra-low endurance, possibly leveraging machine learning for predictive block management. In addition, for latency-sensitive workloads, appropriate hybrid SLC/TLC/QLC modes to improve write performance could be explored.  

\textbf{Read Latency and Error Rate Escalation:}
The tighter voltage threshold distributions in QLC/PLC lead to higher bit error rates and increased read retries \cite{3640372}. This necessitates advanced ECC mechanisms, such as LDPC with soft-decoding and ML-assisted voltage threshold calibration. Besides, Future research could explore low-latency read architectures, potentially leveraging 3D NAND optimizations to minimize access delays.  

\textbf{Data Retention and Temperature Sensitivity:}  
QLC/PLC cells exhibit faster charge leakage, particularly at elevated temperatures, raising data integrity concerns. Potential solutions include temperature-aware refresh algorithms that dynamically adjust retention thresholds, enhanced error recovery schemes for aged data, such as deep learning-based data reconstruction and novel cell materials and structures to improve charge stability in high-density NAND.  

QLC/PLC NAND offers compelling density advantages but faces critical scalability challenges in endurance, latency, and reliability. Addressing these issues requires innovations across device architecture, error correction, FTL design, and materials science. Future research must balance cost, performance, and longevity to enable QLC/PLC adoption in next-generation storage systems.

\subsection{\textbf{Performance vs. Reliability Tradeoff Exploration}}
At the NAND flash level, the trade-off between performance and reliability is most evident in programming technology.
High-density storage modes such as TLC and QLC achieve greater storage capacity by dividing the voltage threshold of storage cells into more discrete levels, significantly reducing the cost per bit, which is a key advantage in the era of data explosion.
However, this comes at the expense of endurance and increased error rates.
Finer charge states increase susceptibility to tunnel oxide damage, accelerate defect accumulation with each programming operation, and significantly reduce P/E cycles.
In addition, the narrower voltage margin between states makes it susceptible to read disturbances and charge leakage, increasing the risk of misreads.

High-density storage modes like TLC and QLC increase capacity by dividing cell voltage thresholds into more discrete levels, reducing cost per bit. However, this sacrifices endurance and increases error rates. Finer charge states accelerate tunnel oxide damage and defect accumulation, reducing P/E cycles. Narrower voltage margins also raise susceptibility to read disturbances and charge leakage, increasing misread risks.

Controller algorithms require balancing reliability and performance: aggressive garbage collection boosts write performance but accelerates wear, while stronger ECC enhances data integrity at the cost of latency and capacity. Future research should develop adaptive garbage collection that dynamically adjusts to real-time usage and wear, and explore efficient ECC techniques to improve integrity while minimizing latency and capacity overhead.

SLC caching uses pseudo-SLC mode to boost write speed but accelerates cell wear, with performance drops occurring during cache flushes. Over-provisioning enhances reliability through spare area allocation, but its efficiency is workload-dependent and degrades over time. Charge leakage, multi-bit error amplification, and operational interference ultimately cause long-term performance-reliability conflicts. Future research should optimize SLC cache management to reduce flush impacts and dynamically adjust over-provisioning allocation based on workload patterns.

Emerging solutions include ML-based wear prediction for dynamic reliability management and namespace partitioning to reduce write amplification via logical-physical data locality coordination. Future work should enhance ML model accuracy and real-time performance for precise wear prediction, and optimize namespace designs to improve locality coordination efficiency for greater write amplification reduction.

\subsection{SSD Optimization for Vector Database in LLM}
Vector databases have become essential infrastructure for high-dimensional similarity search in AI-driven applications. As data volumes grow exponentially, SSD-based vector indexing has emerged as a scalable and cost-effective solution that overcomes the limitations of in-memory systems. However, traditional indexing algorithms face significant efficiency challenges when deployed on SSDs, primarily due to the fundamental mismatch between vector search patterns and SSD storage characteristics. These challenges include excessive random I/O operations, write amplification issues, and high update latency. Although SSDs offer superior bandwidth for large-scale approximate nearest neighbor search, their performance potential is limited by the irregular data access patterns inherent to high-dimensional vector processing.

In terms of SSD-aware index design, Memory-optimized vector indexes (e.g., HNSW, NSG) exhibit poor SSD performance due to random access patterns. DiskANN \cite{jayaram2019diskann} introduces a multi-level grouped layout that replaces random reads with batch sequential operations, aligning with SSD page size and bandwidth to reduce I/O overhead. High-dimensional vector storage demands exceed memory capacity (e.g., 10B vectors require >280TB). 
IVF integration enables hierarchical retrieval with precomputed distances, balancing storage and speed. Traditional layouts ignore vector similarity, causing irrelevant partition scans. Quake \cite{mohoney2025quake} dynamically groups similar vectors via k-means, splitting hot partitions and merging cold ones to concentrate related data, minimize scans, and align storage with access patterns. Vector databases suffer from serial I/O underutilization. Starling \cite{wang2024starling} combines memory navigation graphs with rearranged disk layouts, integrating parallel I/O and computation in block search to maximize SSD bandwidth and reduce latency.

In conclusion, SSD optimization techniques are driving vector databases toward large-scale, low-cost, high-throughput disk-resident architectures. Through collaborative design in indexing, I/O scheduling, and write optimization, researchers and practitioners are unlocking the full potential of SSDs for vector workloads. In the future, as hardware continues to evolve and AI demands intensify, SSD-aware co-processing and energy-efficient designs will become increasingly crucial in vector database systems.

\section{Conclusions}\label{sec:conclusion}
This survey has provided a systematic overview of SSD architectures, key challenges, and optimization techniques at the device level. We began by analyzing the fundamental structure of SSDs, including NAND flash memory operations, controller functionalities (e.g., garbage collection, wear leveling), and host interface protocols, which form the backbone of modern storage systems. We then identified critical challenges, such as reliability degradation due to P/E cycling, endurance limitations in high-density NAND, latency variations, and emerging security threats (e.g., ransomware, data remanence).  

To address these challenges, we explored advanced optimization techniques, including error correction and data reliability enhancements (e.g., LDPC codes, RAID techniques), FTL optimizations (garbage collection, wear leveling, and hybrid designs), discussion of emerging SSD architectures (like OCSSDs, ZNS, and FDP for improved performance and efficiency), and security mechanisms such as secure deletion and plausibly deniable encryption.  

Looking ahead, several open research challenges remain. First, scalability issues with QLC/PLC NAND require innovations in endurance management and error correction. Second, balancing performance and reliability trade-offs, particularly for AI/ML and large-scale data workloads, is important for SSD wide adoption in future applications. Third, future research should optimize SSDs for next-generation applications, including LLM training and KV store.

\bibliographystyle{ACM-Reference-Format}
\bibliography{reference}

\end{document}